\documentstyle[prl,aps,psfig,multicol]{revtex}
\begin{document}
%\preprint{IMSc/96/06/17}
% Title Page
\draft 
\title{ Fermi surface topology and ferromagnetic 
superconductivity in UGe$_2$}
\author{Debanand Sa\cite{e-dsa}}  
%\author{Debanand Sa \footnotemark[1]}
\address{Max Planck Institute for the Physics of Complex Systems, 
N\"othnitzer Str. 38, D-01187 Dresden, Germany}  
%\footnotetext[1]{ E-mail: debanand@iopb.stpbh.soft.net }  
 
%\date{\today}

\maketitle

\begin{abstract}
We consider a Stoner ferromagnet in presence of a quasi-one dimensional 
Fermi surface in its spin majority band. 
Assuming a twin $\delta$-function peaked density of state  
due to the low dimensionality, we computed the single particle 
self-energy. There appears an additional divergence in the self-energy 
and hence in the effective mass inside the ferromagnetic phase (besides 
the standard logarithmic divergence at the Stoner critical point). Since 
such an effect is purely due to density of states, it might correspond   
to a first order phase transition making the ferromagnetic 
phase into two distinct phases. This result is in qualitative agreement 
with the recent specific heat capacity measurement. We also discuss its 
relevance to the superconducting state in UGe$_2$. 
\end{abstract}
%\eject
%\newpage
\begin{multicols}{2}  
\narrowtext 
Superconductivity (SC) and ferromagnetism (FM) are two very different 
cooperative phenomena and the question regarding their coexistence is 
very important theoretically as well as experimentally. Due to their 
antagonistic character, it was believed that they mutually act against 
each other. But the recent discovery of FMSC in UGe$_2$ \cite{sax00}, 
URhGe \cite{aok01} and ZrZn$_2$ \cite{pfl01} has ruled out this possibility. 
The behaviour of these materials is an example of a more general phenomenon 
where a novel state appears on the boarder of magnetism at low temperature. 
In what follows, we concentrate on the material UGe$_2$, where SC is 
entirely covered within the FM phase and disappears in the paramagnetic 
(PM) region. 

UGe$_2$ is an itinerant FM \cite{hux01} with a Curie temperature 
of T$_c$= 52 K at ambient pressure. Upon increasing pressure, it becomes SC 
with a maximum transition temperature of T$_{s}\sim$ 0.8 K in the pressure 
range of 1 GPa to 1.7 GPa.   
Furthermore, there appears an additional phase line T$_x$ \cite{hux01} inside 
the FM phase (see Fig.1) dividing it into two distinct phases (FM1 and FM2). 
This has been inferred from the strong anomaly seen in the resistivity 
measurements \cite{hux01}, a small enhancement in the specific heat capacity 
\cite{tat01}, lattice expansion \cite{oom93} and a change in the character 
of the Fermi surface measured in the de Haas van Alphen experiments 
\cite{ter01}. This might be thought to be associated with the formation of 
charge and spin density waves (CSDW) since the band structure calculations 
\cite{yam93,shi01} indicate that this material is prone to nesting. 
Until now, no density fluctuations has been observed in the neutron 
scattering measurements. 

It is already known that in an itinerant weak FM, the magnetic 
fluctuations are enhanced in the vicinity of 
the quantum phase transition and it gives rise to triplet SC \cite{fay79}. 
The standard phase diagram proposed in such a formalism is the appearance 
of SC dome both in the PM as well as FM phases close to the Curie 
transition. The SC transition temperature estimated in such a formulation 
is comparable in both the phases.  
Recently, similar results have been obtained by solving the Eliasberg 
equations rigorously \cite{rou01}. However, it has also been pointed 
out that SC in the FM region can be enhanced at least 50 times than that 
of the PM phase \cite{kir01}. This can be understood due to the 
mode coupling among transverse magnons and longitudinal spin fluctuations 
in the FM phase which is absent in the PM region. However, the 
FM state of UGe$_2$ is so magnetically anisotropic that the presence of 
magnons seems an unlikely explanation for the enhancement of SC in the 
system. Furthermore, inspired from the band structure calculations 
\cite{yam93,shi01}, Watanabe and Miyake \cite{wat01} postulated the 
existence of CSDW inside the FM phase, in analogy with the 
$\alpha$-phase of uranium. In such a formalism, CSDW fluctuations at 
high wave vector couple to the magnetization in such a way that it 
yields a mass enhancement near CSDW transition. It also provides a 
pairing mechanism for SC in this system.  

%%%%%%%%%%%%%%%%%%%%%%%%%%%%%%%%%%%%%%%%%%%%%%%%%%%%%%%%%%%%%%%%%%%%

\begin{figure} 
\begin{center} 

\psfig{file=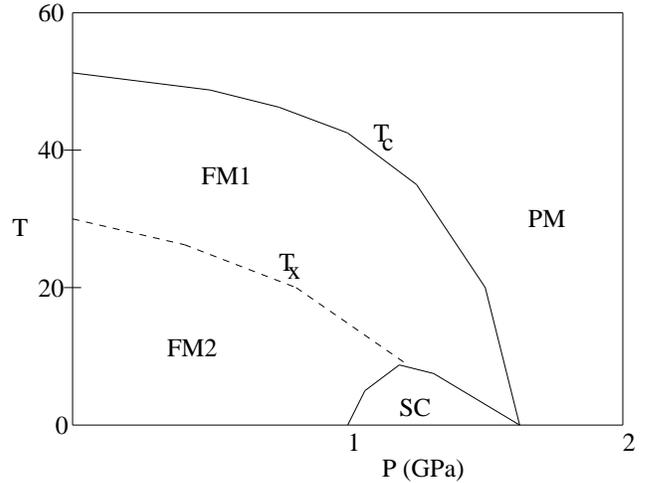}
\medskip  
\caption{\label{phase} Schematic T-P phase diagram in UGe$_2$ 
showing the T$_x$ phase line which separates the two magnetic 
phases FM1 and FM2. } 

\end{center} 
\end{figure} 

%%%%%%%%%%%%%%%%%%%%%%%%%%%%%%%%%%%%%%%%%%%%%%%%%%%%%%%%%%%%%%%%%%%      
\vspace{-0.4cm}
Very recently, the pressure dependence of magnetization measurement at  
low temperature by Pfleiderer et al., \cite{pfl02} indicates that the 
two transitions (Curie as well as the transition between FM1 and FM2) 
are of first order in nature. They further pointed out that even if the 
low temperature uniform 
longitudinal susceptibility undergoes a large change between FM1 and FM2, 
it becomes almost pressure independent within each phase. This suggests 
that the SC is not driven due to its proximity to a FM quantum critical 
point; rather, it could be associated with a sharp spike 
in the electronic density of states (DOS). This could also be 
responsible for the low pressure magnetic transition. Keeping this in mind, 
Sandeman et al. \cite{san02} studied the zero temperature Stoner model 
in presence of a two Lorentzian peaked DOS phenomenologically. 
They minimised the total energy density of the system with respect to the 
magnetization and obtained the magnetization-Stoner exchange (M-U) phase 
diagram. For suitable choices of the DOS parameters, they showed that one 
can obtain two first order transitions. They also discussed its relevance to 
superconductivity. 

In this letter, we study the effect of changing Fermi surface topology 
in an itinerant FM. Due to the hidden quasi-one dimensionality, 
we assume a double $\delta$-function peaked DOS and compute  
the single particle self-energy. Besides the standard logarithmic 
divergence due to spin fluctuations at the Stoner critical point, it yields  
an another divergence inside the FM phase dividing it into two distinct 
phases.  Since this is purely due to DOS effect, one might 
argue this phase line to be of first order in nature. The effective mass 
at this transition gets enhanced, which is in qualitative agreement with 
the recent specific heat capacity measurement. We also discuss the role 
of such DOS in the superconducting phase of UGe$_2$. 

Since the present work is motivated from the material UGe$_2$ where the  
5f electrons from the U-atoms play dual roles both for FM and SC, we 
can start with a model which has Hubbard type exchange interaction.  
This is inferred from the specific heat data \cite{hux01} which is 
at least ten times lower than the standard heavy Fermion materials. 
Further, the measured low residual resistivity ($0.2\>\mu\>\Omega$) 
yields a carrier mean free paths of several thousand angstroms 
\cite{sax00}. Thus, 
the 5f electrons from the U-atoms are considered to be itinerant but  
of course, strongly correlated. We presume that the relevant magnetic 
behaviour of the system is described by Stoner RPA-mean field theory 
\cite{mor}. Further, due to the 
pronounced magnetic anisotropy of UGe$_2$, and as a consequence,  
the absence of tranverse spin modes makes the Stoner approach better at 
low temperature. The energy for single particle excitation in a  
Stoner FM is given as, $\epsilon_{k,\sigma}=\epsilon_k -\mu  
+\sigma\Delta_F $. Here, $\sigma = \pm$ for spin up (down) bands, 
$\Delta_F$ is the order parameter (uniform magnetization) which 
characterizes the itinerant FM state by $2\Delta_F= 
U(<n_{\downarrow}>-<n_{\uparrow}>)$,  
$U$ is the on-site Hubbard interaction energy and $n_{k,\sigma}= 
c_{k,\sigma}^\dagger c_{k,\sigma}$ is the density operator with wave 
vector $k$ and spin projection $\sigma$. The occupation of each spin band 
$\sigma$ is, $n_{\sigma}=\int_{-W/2}^{{\mu}_{\sigma}} 
\rho(\epsilon)d\epsilon$, where $(-W/2)$ is the bottom of the band and spin 
$\sigma$ occupies energy states up to $\mu_{\sigma}$. The DOS here is given 
by $\rho(\epsilon)$. We also consider the total number of spins, 
$N=n_{\uparrow}+n_{\downarrow}$, to be fixed. Thus, the chemical potential 
$\mu_{\sigma}$ of each spin band is completely determined by the particular 
magnetization and the electron number.  

The single particle propagator for a weak FM is given as, 

\begin{equation} 
 G_{\sigma}(k,\omega) = {{Z}\over{\omega-\epsilon_{k,\sigma}+i\delta}} 
+G_{inc}, 
\end{equation} 

\noindent where, the residue $Z$, contains informations about the 
quasiparticle properties (which has to be less than one for well-defined 
quasiparticles as in the framework of Fermi liquid theory). The incoherent 
part of the Green function ($G_{inc}$), which describes the high energy 
processes, is a smooth function of $\omega$ and $k$. In such a situation  
it can only renormalize the properties at the Fermi surface without 
introducing any new physics. Such a liquid has two types of low energy 
collective spin excitations: transverse magnons and longitudinal spin 
fluctuations. For UGe$_2$, due to the large magnetic anisotropy, we  
consider only the latter in what follows. The relevant expression for 
the longitudinal spin fluctuation \cite{izu63,dzy76} is given as, 

\begin{equation} 
\chi_l(q,\omega)= {{\rho(0)}\over{\eta +Aq^2+{{iC\omega}\over{|q|}} }}. 
\end{equation} 

\noindent Here, $\rho(0)$ is the DOS at the Fermi level in the PM state, 
$\eta=(\lambda-1)\sim {{\Delta_F}\over{\epsilon_F}}$, 
$\lambda=U\rho(0)$, A and C are the parameters related to $\epsilon_F$ 
and $\rho(0)$. Now, using the standard many body technique \cite{fet}, 
one can calculate the single particle self-energy in the spin 
majority band as, 

\begin{equation} 
\Sigma(k, \omega)={{\lambda^2}\over{\beta}}\sum_{p, ip_n} 
G^0(k+p, i\omega_n+ip_n)\chi_l(p, ip_n), \label{e} 
\end{equation} 

\noindent  where `$i\omega_n$' represents Matsubara frequencies. At this 
stage of the calculation,  we would like to point out that the wave vector sum 
in the above equation can be performed by using the DOS of the FM state. The 
standard way to get a proper form of the DOS is to analyze the band structure 
data. We do exactly this for the present case. It has already been 
noted from the band structure calculations \cite{yam93,shi01}   
that the Fermi surface in UGe$_2$ is quasi-cylindrical. 
In presence of exchange splitting, the minority electron surface contracts 
while the majority spin counterpart expands towards the boundary of the 
Brillouin zone and gets cut-off by the zone wall on both sides. This 
produces two large and roughly parallel sheets reminiscent of a quasi-one 
dimensional system. It has been known from text books that in 
a perfectly one-dimensional tight binding system, the DOS has square root 
singularity and it diverges at 
both the band edges. One can show that by including the higher order 
harmonics in the dispersion, it is possible to get double peak structure in
the DOS as has been shown by Sandeman et al \cite{san02}.  

To make the $p$-sum tractable analytically in equation \ (\ref{e}), 
we assume a following form  
of DOS, which has double $\delta$-function peaked structure, written as, 

\begin{equation} 
\rho(\epsilon)= \rho(0)[1+a (\delta(\epsilon-b)+\delta(\epsilon+d))].\label{d} 
\end{equation}

\noindent Here, `$a$', `$b$' and `$d$' are the parameters which determine  
the strength and the positions of the DOS. These parameters can be determined 
from the band structure calculations. Moreover, it should be noted that the 
energy $\epsilon$ in the DOS is measured with respect to the Fermi energy of 
the spin majority band ($\epsilon_{F\uparrow}$).  
In what follows, we show that this hidden quasi-one dimensional 
character associated with the large electron sheet of the majority 
spin Fermi surface which yields the above DOS (\ref{d}), will have 
profound consequences. 

Now, by performing the Matsubara frequency summation as well as 
the $p$-sum by using the above form of the DOS, one obtains,   

\begin{equation} 
\Sigma(k_F, \omega)=-{{\lambda^2}\over{4}} \omega\left[{{1}\over{P}}\> 
\ln|1+{{P}\over{\eta}}| 
+ Q\left({{1}\over{\eta+\tilde A b}} 
+ {{1}\over{\eta-\tilde A d}}\right)\right],   
\end{equation} 

\noindent where $P=Ak_F^2$ and $Q=a/\epsilon_F$ are the dimensionless 
variables. Here, $\tilde A=P/\epsilon_F=Ak_F^2/\epsilon_F$. By 
differentiating the real part of the self-energy, one can compute 
the effective mass, which turns out to be, 

\begin{eqnarray} 
\left({{m^*}\over{m}}-1\right) 
=-{{\partial}\over{\partial \omega}} 
{Re\Sigma(k_F,\omega)}|_{\omega=\epsilon_F}  
=g_Z \>\> \nonumber \\
={{\lambda^2}\over{4}} \left[{{1}\over{P}}\> \ln|1+{{P}\over{\eta}}| 
+ Q\>\left({{1}\over{\eta+\tilde A b}} 
+ {{1}\over{\eta-\tilde A d}}\right)\right].\label{m}   
\end{eqnarray} 

\noindent This is the central result in the manuscript. Recalling 
the weak itinerant FM, it is known that the effective mass as well as 
the specific heat capacity diverges logarithmically near the Stoner 
critical point ($\eta\rightarrow 0$). This is exactly the first term 
in the above expression whereas the next two terms are due to the 
double $\delta$-function peaked DOS. It is obvious from the last term 
in the above equation that there is an algebraic divergence for finite 
$\eta$, which causes a phase transition inside the FM phase.
This corresponds to the FM1-FM2 transition which  
has been discussed earlier. The specific heat 
capacity measurement by Tateiwa et al., \cite{tat01} shows that the value 
$C/T$ increases approximately to 95 mJ/mole $K^2$ at pressure about 1.15GPa 
which is nearly the pressure where FM1-FM2 transition occurs. 
This causes an enhancement of the the effective mass as much as three times 
than that at ambient pressure. We can obtain such a mass enhancement from 
the above equation by considering suitable parameters. In the present 
formulation, since the enhancement in the effective mass is with respect 
to the dimensionless coupling parameter ($\eta$), we don't go for a 
quantitative fit 
because the experiments are done with respect to pressure. We would like 
to further make some comments on the second term in equation \ (\ref{m}). 
From the pressure dependence of magnetization, it is clear that even the 
Stoner transition is first order. In a standard weak Stoner FM, the PM-FM 
transition is second order. Thus, 
we can make a choice, $b=0$ in the equation \ (\ref{m}) so that both the 
first and the second term in the above equation cause mass divergence at 
at the same point, $\eta=0$. Thus, one of the peak in the DOS should  
lay exactly at the Fermi level of the spin majority band
($\epsilon_{F\uparrow}$) (see Fig.2).   

%%%%%%%%%%%%%%%%%%%%%%%%%%%%%%%%%%%%%%%%%%%%%%%%%%%%%%%%%%%%%%%%%%%%

\begin{figure} 
\begin{center} 

\psfig{file=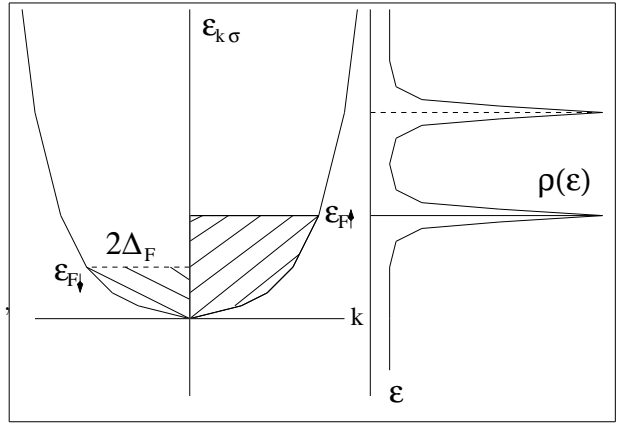}
\medskip  
\caption{\label{dos} Single particle DOS whose one of the peak   
coincides exactly at the Fermi level   
of the spin majority band ($\epsilon_{F\uparrow}$). }

\end{center} 
\end{figure} 

%%%%%%%%%%%%%%%%%%%%%%%%%%%%%%%%%%%%%%%%%%%%%%%%%%%%%%%%%%%%%%%%%%%
\vspace{-0.4cm} 
Next, let us consider the effect of such a twin peak DOS on 
SC. Due to the strong magnetic anisotropy in 
UGe$_2$, it is believed that SC in this system is not only equal spin 
pairing (ESP) but also non unitary \cite{mac01}. Following Fay and Appel 
\cite{fay79}, we can write down the self-consistent gap equation as, 

\begin{eqnarray} 
\Delta_s^{\sigma,\sigma'}(k,i\omega) = {{1}\over{\beta}}\sum_{p,i\nu} 
V^{\sigma,\sigma'}(k,i\omega;p,i\nu)\nonumber \\ 
G_{\sigma}(p,i\nu) G_{\sigma'}(-p,-i\nu) \Delta_s^{\sigma,\sigma'}(p,i\nu),  
\end{eqnarray} 

\noindent where $V^{\sigma,\sigma'}$ is the irreducible pairing potential. 
For a general $l$-state ESP case, one can define a BCS type pairing 
parameter $g_l^{\sigma}$ as, $g_l^{\sigma}=
\rho_{\sigma}(0)V_l^{\sigma,\sigma}$, with 

\begin{equation}
V_l^{\sigma,\sigma}=\int_0^{2k_F}{{qdq}\over{2k_{F\sigma}^2}}
P_l\left(1-{{q^2}\over{2k_{F\sigma}^2}}\right) 
V^{\sigma,\sigma}(|q|,\omega=0).    
\end{equation} 

\noindent Here, $\vec q=\vec k -\vec p$ with 
$|\vec k|\approx|\vec p|\approx k_F$ 
and $(\vec k \cdot \vec p)=k_F^2 \> cos\theta $, $q=2\>k_F\>|sin(\theta/2)|$. 
We can compute the pairing potential $V^{\sigma,\sigma}$ due 
to longitudinal spin fluctuations. For the case of ESP, the ladder diagram 
will be absent due to the Pauli exclusion principle and the contribution 
to $V^{\sigma,\sigma}$ will only be given by summing the odd number of 
bubble diagrams. Within RPA, one obtains, 

\begin{equation} 
V^{\sigma,\sigma} = {{U^2\chi^{-\sigma-\sigma}}\over{1-
U^2\chi^{\sigma\sigma}\chi^{-\sigma-\sigma}}},   
\end{equation} 

\noindent where $\chi^{\sigma\sigma}$ is the particle-hole bubble for a 
particular spin projection $\sigma$. 
Computing  $V^{\sigma,\sigma}$ for a weak itinerant FM as 
has been done earlier \cite{fay79}, one can calculate the BCS parameter 
$g_l^{\sigma}$ by doing angular integration. Moreover, in the present case, 
we can compute $g_1^{\sigma}=g_{1s}$ (we drop the $\sigma$ index since 
we work in a spin majority band, the subscript `$1$'represents the angular 
momentum $l=1$, and `$s$' for the SC state) in presence of a double peaked 
DOS, which turns out to be, 

\begin{equation} 
g_{1s}={{\lambda^2}\over{4}}\left[{{1}\over{P_s}}\> 
\ln|1+{{8\>P_s}\over{\eta}}| 
+Q_s\>\left({{1}\over{\eta}} 
+{{1}\over{\eta-\tilde A_s d}}\right)\right],  \label{s}
\end{equation} 

\noindent where $P_s=\lambda/6 $, $Q_s=2a/5\epsilon_F $ 
and  $\tilde A_s=\lambda/15\epsilon_F $. Now, one can solve the 
self-consistent gap equation within the standard BCS approximation 
which yields, 

\begin{equation} 
k_B \>T_s=1.14\> \omega_c \>\exp[-1/g_{eff}], \label{t} 
\end{equation} 

\noindent where $g_{eff}=g_{1s}/(1+g_Z)$, is the effective coupling and 
$\omega_c$ is the cut-off due to longitudinal spin fluctuations. It has 
been shown earlier by Brinkman and Engelsberg \cite{bri68} that $\omega_c$ 
is in fact proportional to $\eta$. Of course, the derivation of the SC 
transition temperature is admittedly not very satisfactory due to the 
fact that the effective pairing interaction and the self-energy in an 
weakly FM system are energy dependent. So the most elegant method for 
tackling this problem is to go for a conventional Eliasberg strong 
coupling analysis. However, the weak coupling BCS like formulation 
provides a qualitative understanding which is presented above. 

It is clear from equation \ (\ref{m}) and \ (\ref{s}) that the 
effective mass ($g_Z$) as well as the BCS coupling parameter ($g_{1s}$) 
for $l=1$ ESP behaves exactly in the similar way, i.e., they are peaked 
at the Stoner threshold ($\eta=0$) and at the FM1-FM2 transition 
($\eta\sim \tilde A_s d$). Furthermore, the effective coupling parameter 
($g_{eff}$) also has similar feature as that of $g_Z$ and $g_{1s}$ but 
the absolute magnitudes are different. This indicates that the factor 
$\exp[-1/g_{eff}]$ also attains a maximum at $\eta=0$ and 
$\eta\sim \tilde A_s d$. 
However, due to the presence of the cut-off as a prefactor in the expression 
for $k_B \>T_s$, the SC transition temperature vanishes at the Stoner 
threshold while it becomes maximum near the FM1-FM2 transition. This 
is what is exactly observed in the material UGe$_2$.  Thus, as far as 
the transport and the SC is concerned, the present formulation qualitatively 
explains the experimentally observed spectrum.         

To conclude, we summarize the main results of the present work. 
We study the role of hidden quasi-one dimensional Fermi surface in the 
spin majority band of a weak Stoner FM. This was inferred from the band 
structure calculations in the material UGe$_2$. To make the calculations 
analytically tractable, we assume a double $\delta$-function peaked DOS 
and computed the single particle self-energy. We showed that the effective 
mass and hence the specific heat capacity acquires an additional divergence 
besides its standard logarithmic divergence at the Stoner critical point. 
This ultimately divides the FM phase into two distinct phases.  
Since this is purely due to the DOS effect, one might 
argue this transition to be of first order in nature. 
It is also shown that such a DOS yields maximum SC transition temperature 
at a point at which the specific heat capacity gets enhanced. This is in 
qualitative agreement with the phase diagram obtained for UGe$_2$. 

%\bigskip   
The author would like to thank R. Narayanan and Sharon Sessions for 
useful discussions and carefully reading the manuscript.  
%\newpage

%\begin{references}
 
%\end{references}  

\end{multicols}  
\end{document}